\documentclass[12pt,onecolumn,draftclsnofoot]{IEEEtran}
\usepackage{amsmath,amsfonts,amssymb,mathrsfs,amsthm}
\usepackage{latexsym}
\usepackage{stmaryrd}
\usepackage{latexsym}
\usepackage{epsfig}
\usepackage{hyperref}
\usepackage{epstopdf}
\usepackage[nospace,noadjust]{cite}
\usepackage{array}
\usepackage[english]{babel}
\usepackage{color}
\usepackage{babel}
\usepackage{graphicx,xcolor,colortbl}    
\usepackage{epstopdf,epsfig}
\usepackage{graphicx}
\usepackage{tabularx,ragged2e,booktabs,caption}
\usepackage{algorithm,algpseudocode}
\usepackage{tikz}
\usepackage{graphicx}
\usepackage{caption}
\usepackage{hhline}
\usepackage{caption}
\usepackage{subcaption}
\usepackage{multirow}
\usetikzlibrary{shapes,arrows}
\usepackage{xcolor,colortbl}

\newtheorem{lemma}{\noindent {\bf Lemma}}

\include{new_commands}

\makeatletter
\let\l@ENGLISH\l@english
\makeatother

\makeatother

\begin{document}%
\title{\LARGE A QoS-Oriented Trajectory Optimization in Swarming
Unmanned-Aerial-Vehicles Communications}
\author{
Amine Bejaoui,~\IEEEmembership{Student Member,~IEEE,} Ki-Hong Park,~\IEEEmembership{Member,~IEEE}, and Mohamed-Slim Alouini,~\IEEEmembership{Fellow,~IEEE}

\vspace{-0.3in}
}

\maketitle

\begin{abstract}

This letter aims to present a novel approach for unmanned aerial vehicles (UAV)' path planning with respect to certain quality of service requirements. More specifically, we  study  the max-min fairness problem in an air-to-ground communication system where multiple UAVs and multiple ground stations exist. We jointly optimize the UAVs’ trajectories and power allocation as well as the user scheduling. To this end, we propose an effective iterative algorithm that relies on the successive convex approximation  and the block coordinate descent techniques.
\end{abstract}
\section{Introduction}
Recently, unmanned aerial vehicles (UAVs) 
have drawn significant research interests \cite{27,28}.  This is mainly due to the significant decrease in the production costs of these systems \cite{a10}, added to their flexible deployment and high maneuverability. Traditionally, the use of UAVs was only dedicated  to military applications in the late 90’s~\cite{a3}. 
However, research and development of UAVs have 
enabled the emergence of several new applications. As a matter of fact, the drones have started to be used in cargo delivery, traffic control, weather forecasting, farming \cite{a9}, etc. 

Motivated by this context, earlier research started by studying  single UAV systems. To begin with, they studied the problem of path planning that consists into scheduling the areas to visit in a given time period. For instance, the authors in \cite{1} presented a multi-colony algorithm that optimizes the UAV path planning with obstacle avoidance.

After that, recent research has investigated the use of multi-UAV systems. In fact, it has been proved that, compared to a single UAV system, the use of a fleet of UAVs allows to perform more complex and challenging missions with higher speed and especially with increased performance and efficiency. However, with the use of a multi-UAV system, we are facing a new kind of problems such as interference, energy efficiency, path planning and user scheduling. For instance, the work in \cite{24} has treated the path planning problem for a multi-UAV system using genetic algorithms and Bezier curves. Furthermore, a game theoretic approach has been presented in \cite{26} to reduce the interference between the UAVs. However,  there is only the work in \cite{a22} that considered working on all these problems combined. However, in that work, every single one of these problem was optimized separately. 
In the same context, we will conduct different optimization approach to get better fairness among served users.

In this letter, we study a multi-UAV aided wireless system where a group of UAVs are used to provide down-link communication to a group of ground stations (GT) in a two dimensional area given certain quality of service requirements in communication. The objective here is to maximize the minimum average rate among ground stations in order to achieve fair performance among them during a given finite period. To this extent, we consider a joint optimization problem in which we have to provide an efficient flight trajectory solution for every UAV with specifying the optimal UAV-user association and power allocation at each time. To that end, we propose an iterative algorithm based on the 
 successive convex approximation  (SCA) and the block coordinate descent (BCD) techniques. The proposed algorithm and the resulting fairness will be verified by simulation results.
\section{System model and problem formulation}
\label{System}
We consider a wireless communication system where $M$ UAVs are employed to communicate with $K$ ground stations. 
Furthermore, we assume  that every ground station $k\; (k=1,..,K)$ has a fixed position $\mathbf{w}_{k}=[x_{k},\;y_{k}]$ on the ground, while all the UAVs $(m=1,..,M)$ fly on the two dimensional coordinate $\mathbf{q}_{m}(n)=[x_{m}(n),\;y_{m}(n)]$ at a fixed altitude $H$\footnote{We suppose here that the trajectory optimization problem is intractable over continuous time. For ease of exposition,  we use  the  discrete  linear  state-space  approximation. This method consists of discretizing the time horizon $T$ into $N$ time slots with a step size  $\delta t$. So, we have  $t = n\delta t$ with  $n = 0,1,\ldots , N$. Thus all the formulas in this letter will be defined as a function of the discrete time instant $n$.}. 
Thus, the distance between an UAV $m$ and a ground station $k$ is expressed by $d_{k,m}(n)= \sqrt{H^2+ \| \mathbf{q}_{m}(n)- \mathbf{w}_{k} \| ^2} $. For ease of exposition, we assume that the high altitude of UAVs enables them to effectively establish line-of-sight (LoS) link and the Doppler effect caused by their mobility is well compensated at the receivers. Therefore, the time-varying channel between an UAV $m$ and a ground station $k$ follows the free-space path loss model, which can be expressed as $h_{k,m}(n)= \frac{\beta_{0}}{H^2+ \| \mathbf{q}_{m}(n)- \mathbf{w}_{k} \| ^2} $, where $\beta_{0}$ represents the channel gain at reference distance ($d=1$ m).
\par 
We also assume  that all the UAVs operate in the same frequency band. 
Thus, the corresponding received signal to interference plus noise ratio (SINR) at user $k$ from UAV $m$ is  computed as 
\begin{eqnarray}
\gamma_{k,m}(n)= \frac{p_{m} (n)  h_{k,m}(n)}{\sum_{\substack{
   j=1 \\
   j\neq m
  }} ^{M}
 p_{j} (n)  h_{k,j}(n)+ \sigma_{0} ^2} ,
\end{eqnarray}
where $ p_{m}(n)$ denotes the down-link transmit power of UAV $m$, $\sigma_{0}^2$ represents the variance of additive  white Gaussian noise (AWGN) at the receiver  and the term $\sum_{\substack{
   j=1 \\
   j\neq m
  }} ^{M}
 p_{j} (n)  h_{k,j}(n)$ stands for the co-channel interference (CCI) received by ground station $k$ during its communication with the UAV $m$ at time $n$. 
We consider the power control of the signals transmitted by  UAVs to  ground stations at each time $n$ in order to manage co-channel interference to the other users. 
The down-link transmit power of an UAV $m$ at time instant $n$ is constrained on $0\leq p_{m} (n)  \leq P_{max}$, where $P_{max} $ denotes the maximum allowable transmit power at the UAV.
\par 
In order to define the UAV-user association and scheduling, we use a binary variable $\alpha_{k,m}(n)$ that takes 1 during communication between an UAV $m$ and a ground station $k$ in time $n$. Otherwise, it is equal to 0. Then, we can  define the achievable rate of user $k$ over the period $N$ is given by:
\begin{eqnarray}
R_{k}=\frac{1}{N}  \displaystyle\sum_{n=0}^{N} \sum_{m=1}^{M}\alpha_{k,m}(n)\log_{2}(1+\gamma_{k,m}(n)).
\end{eqnarray}

Each UAV operates with a finite amount of 
energy due to the compact size requirement which is limiting the serving time and data rate on flight. As explained in \cite{123},  we assume that the communication energy consumption is much smaller than the propulsion energy consumption and we only consider the propulsion power as the source of energy consumption, which can be  expressed by
\begin{eqnarray} 
E_{m}&=& \displaystyle\sum_{n=0}^{N} c_{1} \| \mathbf{v}_{m} (n) \| ^3+\frac{c_{2}}{\| \mathbf{v}_{m} (n) \| } \left(1+\frac{\| \mathbf{a}_{m} (n) \|^2}{g }\right)  \nonumber\\
&+&\frac{1}{2}w (\| \mathbf{v}_{m} (N) \|^2-\| \mathbf{v}_{m} (0) \|^2)\leq E_{max}, \indent \forall m, \label{constEm}
\end{eqnarray}
where $c_{1}$ and $c_{2}$ are constants, $g$ is the gravitational acceleration, $w$ is the mass of an UAV and $E_{max}$ is the maximum allowable propulsion energy during the period $N$.  In (\ref{constEm}) $\mathbf{v}_m(n)$ and $\mathbf{a}_m(n)$ are the velocity and acceleration of UAV $m$ at time instant $n$ which are approximated by the first order Taylor expansion as
\begin{eqnarray}
\mathbf{v}_{m}(n+1) &=& \mathbf{v}_{m}(n) + \mathbf{a}_{m}(n) \delta t, \label{a5}\\
\mathbf{q}_{m}(n+1) &=& \mathbf{q}_{m}(n) + \mathbf{v}_{m}(n) \delta t + \frac{1}{2}\mathbf{a}_{m}(n) \delta  t^2.\label{a6}
\end{eqnarray}
The velocity and acceleration speed are limited to
\begin{eqnarray}
\| \mathbf{v}_{m}(n) \| &\leq& V_{max} ,\hspace{0.5cm}\forall m,n,  \label{vmax} \\
\| \mathbf{a}_{m}(n) \| &\leq& a_{max} ,\hspace{0.5cm}\forall m,n , \label{amax}\\
\| \mathbf{v}_{m}(n) \|  & \geq& V_{min} ,\hspace{0.5cm}\forall m,n, \label{a3}  
\end{eqnarray}
where $V_{max}$ and $V_{min}$ are the maximum and minimum velocity during flight and $a_{max}$ is the maximum acceleration.

In order to avoid collusion, we add a constraint that sets a minimum distance between the UAVs which is defined as
\begin{eqnarray}
\| \mathbf{q}_{m}(n)-\mathbf{q}_{j}(n) \| &\geq& d_{min} ,\hspace{0.5cm}\forall m,n, j\neq m  \label{dmin} 
\end{eqnarray}

We suppose  that, at each time slot, each UAV serves at most one ground station and that each ground station is served at most by one UAV. Thus, the constraints on association binary variables are given by
\begin{eqnarray}
\displaystyle\sum_{k=1}^{K} \alpha_{k,m}(n) \leq 1,   \indent   
\displaystyle\sum_{m=1}^{M} \alpha_{k,m}(n) \leq 1,   \indent \forall k,m,n. \label{ainf}
\end{eqnarray}

\section{Max-Min Rate Problem Formulation}
In order to better understand the problem, we adopt the following notations :
\begin{itemize}
    \item  $\mathcal{S}$= $\{ \alpha_{k,m}(n)$ \hspace{0.1 cm} $\forall k,m,n\}$: the user-UAV association.
    \item $\mathcal{P}$= $\{p_{m}(n)$, \hspace{0.1 cm} $\forall m,n\}$: the transmit power of the UAVs.
    \item $\mathcal{Q}$= $\{\mathbf{q}_{k,m}(n)$, \hspace{0.085 cm} $\forall k,m,n\}$: the UAVs' trajectories.
    \item $\mathcal{V}$= $\{\mathbf{v}_{k,m}(n)$, \hspace{0.1 cm} $\forall  k,m,n\}$: the UAVs' velocities.
    \item $\mathcal{A}$= $\{\mathbf{a}_{k,m}(n)$, \hspace{0.1 cm} $\forall k,m,n\}$: the UAVs' accelerations.
\end{itemize}
\par
As explained in the first section, we want to jointly optimize the down-link communication and the UAV-user scheduling as well as the trajectory, velocity and acceleration of each UAV in order to maximize the minimum average rate among all users. To this end, if we set $\mu= \min\limits_{k \in \mathcal{K}} R_{k}$, we end up with the following problem \textbf{(P1)} :

  \begin{eqnarray}
&\max\limits_{\mu,\mathcal{Q,V,A,S,P}}&  \mu \nonumber   \\
&\textrm{s.t.} & (\ref{constEm})-(\ref{ainf}) \nonumber \\
(\textbf{P1})&&  R_{k} \geq \mu ,\indent\forall k  \label{a1}  \\
&& 0\leq p_{m} (n)   \leq P_{max}  ,\indent\forall m,n,  \label{afin} \\
&&\alpha_{k,m}(n)\in \{0,1\},\indent\forall k,m,n, \label{a2}
\end{eqnarray}
Problem  (\textbf{P1}) is challenging to solve due to the following reasons. To begin with, 
our problem includes integer constraint that is expressed in (\ref{a2}) due to the presence of the binary optimization variables of UAV-user scheduling and association. Furthermore, 
the constraints (\ref{constEm}), (\ref{a3}), (\ref{dmin}), and (\ref{a1})  are not convex with respect to either $\mathcal{Q,V,A,P}$. Finally, all the optimization variables are closely correlated. For instance, even a slight change of a given UAV trajectory can have an enormous impact on the trajectories of the other UAVs. So, we need to jointly optimize these variables in order to reach the better solution. All these constraints yield  a mixed-integer non-convex problem and, in general, there is no standard method for solving this kind of problems efficiently.
 
\section{Proposed Algorithm}
Our problem is a mixed-integer  non-convex  problem that involves integer constraints. To tackle our problem of interest, we  relax our problem by changing the binary variables of UAV-user scheduling into continuous variables. So, the resulting problem \hypertarget{P2}{(\textbf{P2})} is:
\begin{eqnarray}\label{P2}
&\max\limits_{\mu,\mathcal{Q,V,A,S,P}}&  \mu  \nonumber   \\
(\textbf{P2})&\textrm{s.t.}&   (\ref{constEm})-(\ref{afin}) \nonumber   \\
&&  0\leq \alpha_{k,m}(n)\leq 1,\indent\forall k,m,n.  \label{arelax}
\end{eqnarray}
\par
This relaxation implies that the objective value of the new problem (\textbf{P2}) will be considered as an upper bound of the problem (\textbf{P1}). Thus, from now on, we will optimize this new problem (\textbf{P2}) which is still non-convex due to the constraints (\ref{constEm}), (\ref{a3}) and (\ref{a1}).
\par
In order to do so, we use an iterative algorithm by leveraging the BCD method. Specifically, we divide our optimization variables into two blocks; the first block is for the user scheduling and association ($\mathcal{S}$) and the second one is represented by the UAVs trajectory and power control ($\mathcal{Q,V,A,P}$). So, these two blocks are alternately optimized in each iteration. 
As a result, we obtain two sub-problems of problem (\textbf{P2}).

In the first sub-problem, we only retain the constraints that are related to the user scheduling, while other variables ($\mathcal{Q,V,A,P}$) are fixed
. Thus the first sub-problem \hypertarget{P2.1}{(\textbf{P2.1})} can be expressed as: 
\begin{eqnarray}
(\textbf{P2.1}) &\max\limits_{\mu,\mathcal{S}}&  \mu \indent \textrm{\textrm{s.t.}.} \indent (\ref{ainf}), (\ref{a1}) \textrm{ and }  (\ref{arelax}).\nonumber
\end{eqnarray} 
(\textbf{P2.1}) is a standard linear problem. So, it can be solved directly by using some known optimization tools. 
\par
In the second sub-problem \hypertarget{P2.2}{(\textbf{P2.2})}, we only retain the constraints that depend on the UAVs' trajectory, velocity, acceleration and power control. Thus, (\textbf{P2.2}) is defined by 
\begin{eqnarray}
(\textbf{P2.2}) &\max\limits_{\mu,\mathcal{Q,V,A,P}}&  \mu  \indent \textrm{\textrm{s.t.}.} \indent (\ref{constEm})-(\ref{dmin}),\; (\ref{a1}),\; (\ref{afin}). \nonumber
\end{eqnarray} 
The sub-problem (\textbf{P2.2}) is still non-convex due to the constraints (\ref{constEm}), (\ref{a3}), (\ref{dmin}) and (\ref{a1}). In order to solve it, we first couple the trajectory and the transmit power variables by introducing an auxiliary variable $B_{k,m}(n)$ which is defined by  $B_{k,m}(n) =p_{m} (n)  h_{k,m}(n)$. As a result, $R_{k}$ is rewritten as
\begin{eqnarray}
   R_{k}=  \frac{1}{N} \displaystyle\sum_{n=0}^{N} \displaystyle\sum_{m=1}^{M}\alpha_{k,m}(n) \log_{2}\left(1+ \frac{B_{k,m}(n)}{\sum_{\substack{
   j=1 \\
   j\neq m
  }} ^{M}
 B_{k,j}(n)+ \sigma_{0} ^2}{}\right).\nonumber
\end{eqnarray} 
 \par
 Additionally, the constraint (\ref{afin})  is transformed into
 \begin{eqnarray}\label{a15}
    0\leq B_{k,m} (n)  \leq \frac{\beta_{0}P_{max}}{H^2+ \| \mathbf{q}_{m}(n) - \mathbf{w}_{k}\| ^2}{} \indent  \forall k,m,n, \end{eqnarray}
which is still non-convex. Thus, if we set $\mathcal{B}$= [$B_{k,m}(n),$ $\forall k,m,n$], we have the new problem \hypertarget{P3}{(\textbf{P3})} which can be formulated as

\begin{eqnarray}
(\textbf{P3}) &\max\limits_{\mu,\mathcal{B,Q,V,A}}&  \mu \indent \textrm{s.t.} \indent  (\ref{constEm})-(\ref{dmin}), \;(\ref{a1}),\; (\ref{a15}).\nonumber    
\end{eqnarray} 

In the following, we will try to solve the sub-problem (\textbf{P3}) using the successive convex optimization techniques. The main idea consists in finding another convex problem which represents a lower bound of (\textbf{P3}). Then, we successively maximize the lower bound of (\textbf{P3}) via optimizing the trajectory and the auxiliary variable.

Using the approximations in \hyperlink{Appendix A}{\textbf{(Appendix A)}}, we can find a convex problem \textbf{(P4)} for a given trajectory, velocity and the auxiliary variable ($\mathcal{Q}^r,\mathcal{V}^r,\mathcal{B}^r$) at iteration $r$ defined by:
\begin{small}
\begin{eqnarray}
\hypertarget{P4}{}
& \max\limits_{\mu^{lb},\mathcal{B,Q,V,A},\Lambda} & \mu^{lb} \nonumber\\ 
& \textrm{s.t.} & (\ref{a5})-(\ref{amax}) \nonumber\\
&&  \frac{1}{N} \displaystyle\sum_{n=0}^{N} \displaystyle\sum_{m=1}^{M}\alpha_{k,m}(n)\log_{2}\left(\sum_{\substack{
   j=1
  }} ^{M}B_{k,j}(n)+\sigma_{0}^2\right) - \overline{R}_{k,m}^{upper}(n))\geq \mu^{lb}, \indent  \forall k \nonumber\\
&(\textbf{P4})& 0\leq B_{k,m}(n)\leq  P_{max}\beta_{0}\left(- \frac{\| \mathbf{q}_{m}(n) - \mathbf{w}_{k}\| ^2}{H^4}+D_{k,m}(n)(\mathbf{q}_{m} (n) - \mathbf{w}_{k})^T(\mathbf{q}_{m}^r (n) - \mathbf{w}_{k})+F_{k,m}(n)\right), \forall k,m,n  \nonumber\\
&& 0\leq \displaystyle\sum_{n=1}^{N} c_{1} \| \mathbf{v}_{m}(n) \| ^3  +\frac{c_{2}}{\lambda_{m}(n) }\left(1+\frac{\| \mathbf{a}_{m}(n) \|^2}{g }\right)+\frac{w}{2} (\| \mathbf{v}_{m}(N) \|^2-\| \mathbf{v}_{m}(0) \|^2)    \leq E_{max},  \indent \forall m \nonumber \\
&& \lambda_{m}(n)\geq V_{min}, \indent \forall m,n \nonumber \\
&& \lambda_{m}(n) ^2 \leq \| \mathbf{v}_{m}^r (n) \|^2 + 2(\mathbf{v}_{m}^{r}(n))^T\left( \mathbf{v}_{m}(n)-\mathbf{v}_{m}^r (n)\right),	\indent \forall m,n  \nonumber \\
&& \|\mathbf{q}_{m}(n)-\mathbf{q}_{j}(n) \|^2 \leq \| \mathbf{q}_{m}^r (n)-\mathbf{q}_{j}^r (n) \|^2 + 2(\mathbf{q}_{m}^r (n)-\mathbf{q}_{j}^r (n))^T\left( \mathbf{q}_{m}^r (n)-\mathbf{q}_{j} (n)-\mathbf{q}_{m} (n)+\mathbf{q}_{j}^r (n)\right),	\indent \forall m,n,j\neq m  \nonumber 
\end{eqnarray}
\end{small}
where  $\Lambda=\{\lambda_{m}(n)$ \hspace{0.1 cm} ,$\forall m,n$\} is a set of slack variables that we introduced to the problem, and the constants $D_{k,m}(n)$, $F_{k,m}(n)$ and  $\overline{R}_{k,m}^{upper}(n)$  are given by:
\begin{small} \begin{eqnarray}
    D_{k,m}(n)\!\!&\!\!=\!\!&\!\!2\left(\frac{1}{H^4}-\frac{1}{(\| \mathbf{q}_{m}^r (n) - \mathbf{w}_{k} \| ^2+H^2)^2}\right)\nonumber\\
    F_{k,m}(n)\!\!&\!\!=\!\!&\!\!\frac{1}{\|\mathbf{q}_{m}^r (n) - \mathbf{w}_{k} \| ^2+H^2}+ 2 \frac{1\| \mathbf{q}_{m}^r (n) - \mathbf{w}_{k} \| ^2}{(\| \mathbf{q}_{m}^r (n) - \mathbf{w}_{k} \| ^2+H^2)^2}\nonumber
    - \frac{\| \mathbf{q}_{m}^r (n) - \mathbf{w}_{k} \| ^2}{H^4} \nonumber\\
    \overline{R}_{k,m}^{upper}(n)\!\!&\!\!=\!\!&\!\!\sum_{\substack{
   j=1 \\
   j\neq m
  }} ^{M}
\left(\frac{\log_{2}(e)}{\sum_{\substack{
   j=1\\ j\neq m
  }} ^{M}B_{k,j}^r(n)+\sigma_{0}^2}\right)\left(B_{k,j}(n)-B_{k,j}^r(n)\right)\nonumber+ \log_{2}\left(\sum_{\substack{
   j=1\\ j\neq m
  }} ^{M}B_{k,j}^r(n)+\sigma_{0}^2\right). \nonumber
\end{eqnarray} 
\end{small}

It is worthy noting that the feasible set of problem (\textbf{P4}) can be considered as a subset of the sub-problem (\textbf{P2.2}). 
Thus, by solving iteratively the problem (\textbf{P4}), we will be successively optimizing the lower bound of sub-problem (\textbf{P2.2}) in order to better approximate it. Thus, we will end up with an efficient approximate solution of the latter.
Finally, we present an overall algorithm that synthesizes all the work done in this perspective:
\vspace{1mm}
\hrule
\vspace{2mm}
\textbf{\textbf{Algorithm :} (BCD) Algorithm for Problem (\textbf{P2}). }  \vspace{2mm}

\hrule
\vspace{2mm}
\textbf{Input} : Locations of GTs, $V_{max},V_{min},A_{max},P_{max}, $ $E_{max}, \sigma_0^2, \beta_0, c_1, c_2$ and $\epsilon$
\begin{enumerate}
  \item Initialize $\mathcal{Q}^0,\mathcal{V}^0,\mathcal{B}^0$ and $r=0$
  \item \textbf{Repeat}
  \begin{enumerate}
 \item Solve problem (\textbf{P2.1}) for given $\mathcal{Q}^r$,$\mathcal{V}^r$,$\mathcal{B}^r$, and denote the
optimal solution as $\mathcal{S}^{r+1}$.
 \item \hypertarget{2.b}{}Solve problem (\textbf{P4}) for given $\mathcal{S}^{r+1}$, and denote the
optimal solutions as $\mathcal{Q}^{r+1}$,$\mathcal{V}^{r+1}$,$\mathcal{B}^{r+1}$.
\item Update $r= r+1$.
\end{enumerate}
\item \textbf{Until} \hspace{0.1 cm} The fractional increase of the objective value of problem (\textbf{P2}) is below a threshold $\epsilon$ .
\item Define  \hspace{0.1 cm} $p_{m}(n)=\displaystyle\sum_{k=1}^{K}\alpha_{k,m}(n) \frac{B_{k,m}(n)}{h_{k,m}(n)}, \;\;\; \forall m,n$.
\end{enumerate}
\hrule
\vspace{0.1in}
As explained in \hyperlink{Appendix B}{\textbf{(Appendix B)}}, our algorithm performs better at each iteration and is guaranteed to converge. Also, the complexity of our algorithm is of order {\small{ $\Theta\left((KM)^{3}N +(KMN)^{2}(KMN+\frac{M^2N}{2})\right)$. }} But due to space limitation,  the proof of convergence and complexity are shown in the companion report \cite{l}.  Also, it has been proved in \cite{com} that the convergence complexity of our algorithm in terms of number of iterations is  $\Theta (1/r)$ (i.e.$f\left(x^{r}\right)-f^{*}= \Theta\left(\frac{1}{r}\right)$, where $f^{*}$ is the optimal objective value and $f\left(x^{r}\right)$ is the objective value at iteration r). Thus, this indicates that our algorithm converges in a few iterations.  We will prove that our algorithm performs very well by testing its effectiveness on some examples.
\section{Numerical results}
\par
We consider  a system model composed of $M=2$ UAVs at a fixed altitude $H= 100$ m and $K=6$ GTs on the ground that are arbitrarily distributed in an area of $500\times500$ m$^2$ for simulation purpose. For channel gain, we set $P_{max} = 0.1$ Watt, $\beta_{0} = -60$ dB  and $\sigma_{0}^2 = -110$ dBm, respectively. For propulsion energy and flight dynamics, we set  $V_{min}=1.5$ m/s, $V_{max}=50 $ m/s, $a_{max}=5$ m/s$^{2}$, $E_{max}=2\times 10^{5} J$, $c_{1} = 9.26 \times 10^{−4}$ and $c_{2} = 2250$, respectively. Simulation period is $N=100$ with step size $\delta t=1$ sec.

Next, in order to use our algorithm, we need to provide an initial scheme for the different variables of our model. To that end, we start by providing a simple systematic circular trajectory   design for all the UAVs using the $K$-mean clustering algorithm to find the radius and center (also called centroid) of each trajectory as explained in \hyperlink{Appendix C}{\textbf{(Appendix C)}}. We also assume that the initial speed of every UAV $m$ is set to be constant during all the flight  which are respectively equal to $3$ and $4 m/s$.

Regarding the user scheduling, at each time slot, every UAV $m$ will communicate with the closest ground station that belongs to its assigned cluster $m$ that results from the $K$-mean clustering algorithm. Also, the transmission power of all the UAVs is set to be at its maximum level $P_{max}$.
\par 
Thus, we end up with an efficient initialization scheme for our problem due to the following reasons. First, this method offers a feasible solution that is compliant with all the constraints of our problem. Second, by using this method, we ensure that all the UAVs will provide full coverage for the areas that contains the ground stations. As a result, all the ground stations will be served equally during the flight. Thus, this will help  achieve fair performance between all the ground stations which represents our main goal from the beginning.
  
 \par Finally, since the $K$-mean clustering algorithm divides the ground stations into clusters that are generally far away from each other, we can guarantee that the UAVs are sufficiently separated during their communications with the ground stations. As a result, this tends to minimize the co-channel interference on the ground stations and thus, the achievable data rate at the ground stations will be at its maximum level.


\begin{figure}[H]
\begin{subfigure}{.5\textwidth}
  \centering
  \includegraphics[width=\linewidth]{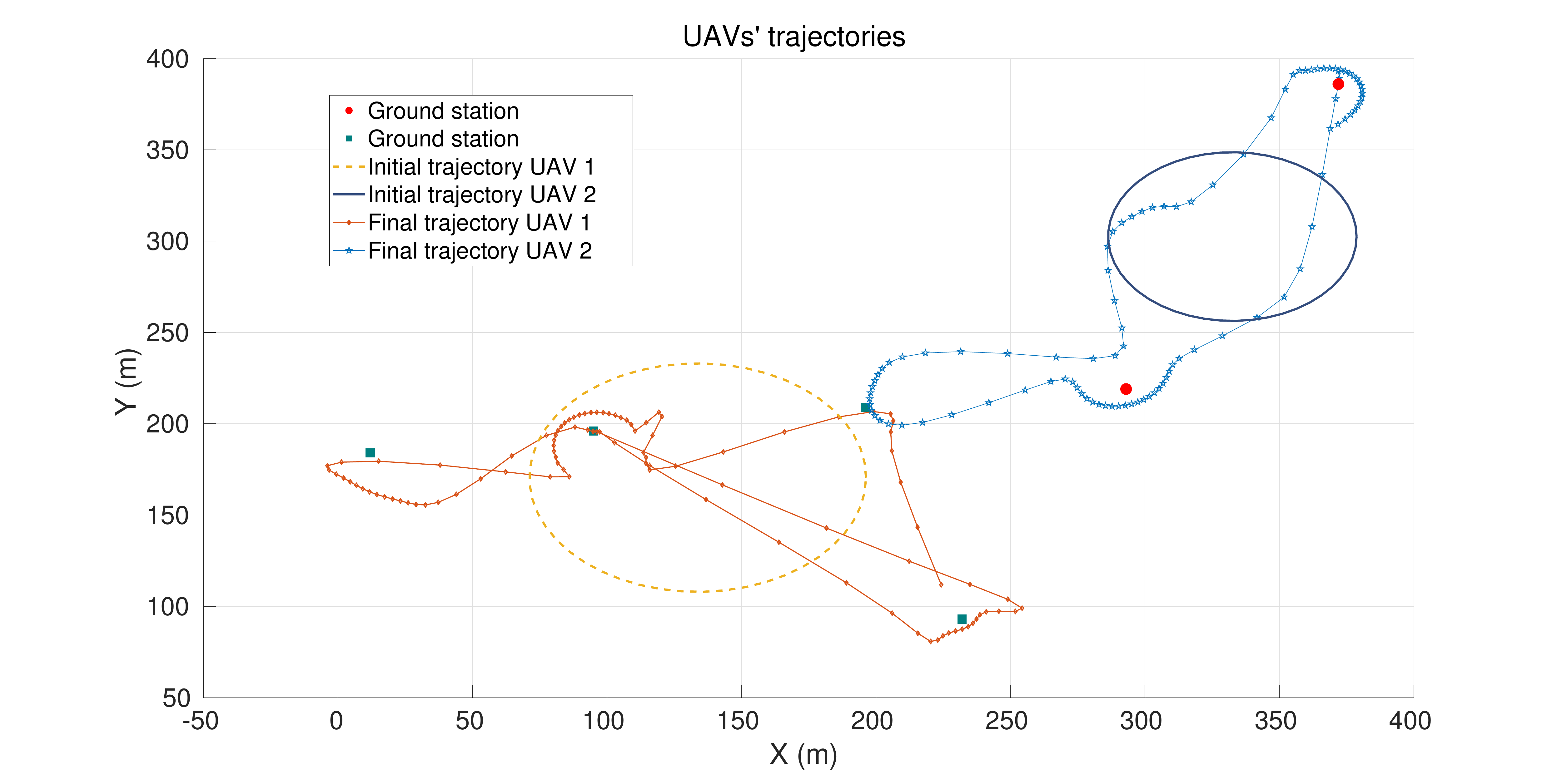}  
  \caption{UAVs' trajectory for $T=100$ s}
  \label{fig:sub-first}
\end{subfigure}
\begin{subfigure}{.5\textwidth}
  \centering
  \includegraphics[width=\linewidth]{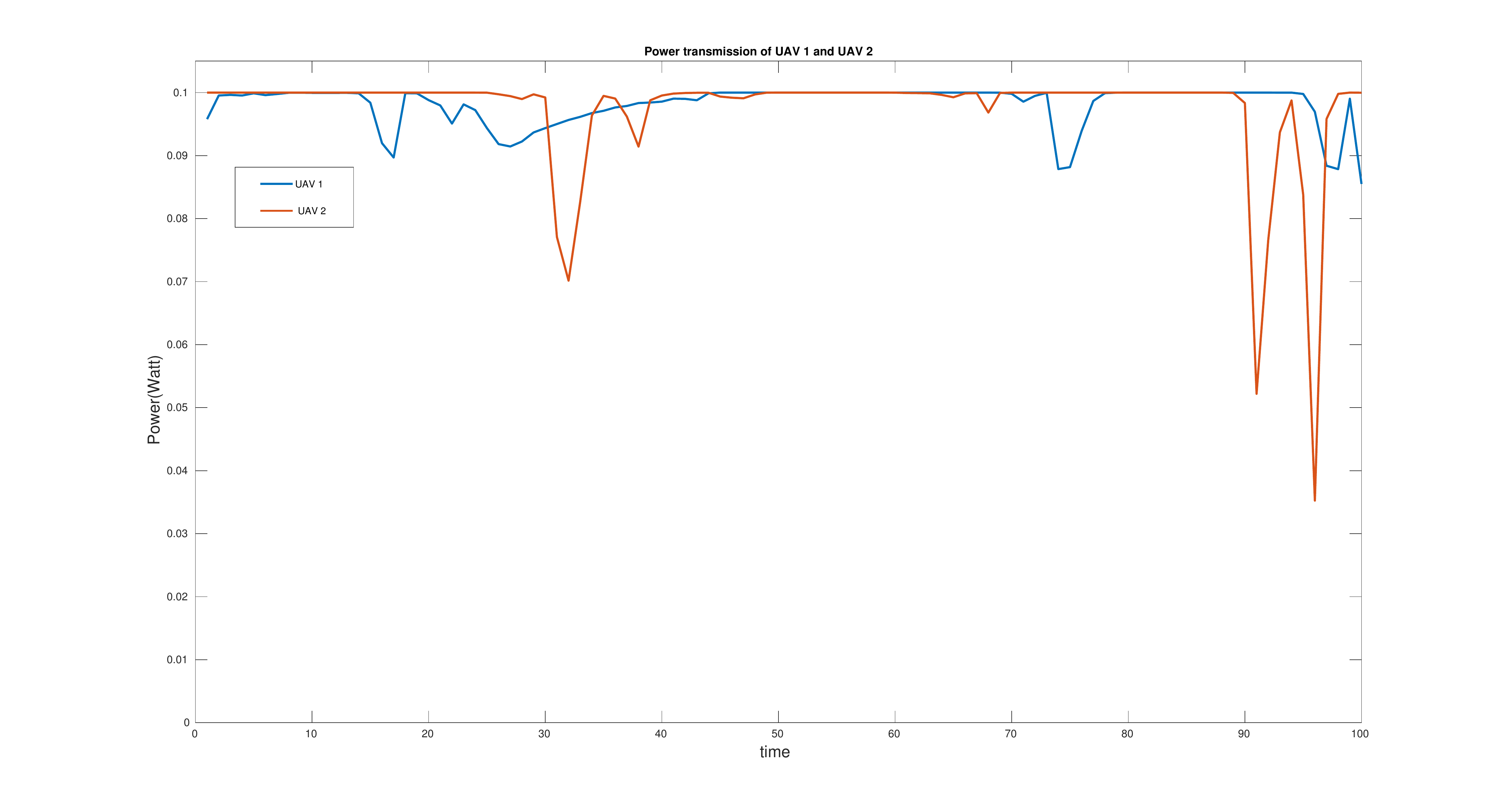}  
  \caption{UAVs' power transmission for $T=100$ s}
  \label{fig:sub-second}
\end{subfigure}
\newline
\begin{subfigure}{\textwidth}
  \centering
  \includegraphics[width=.8\linewidth]{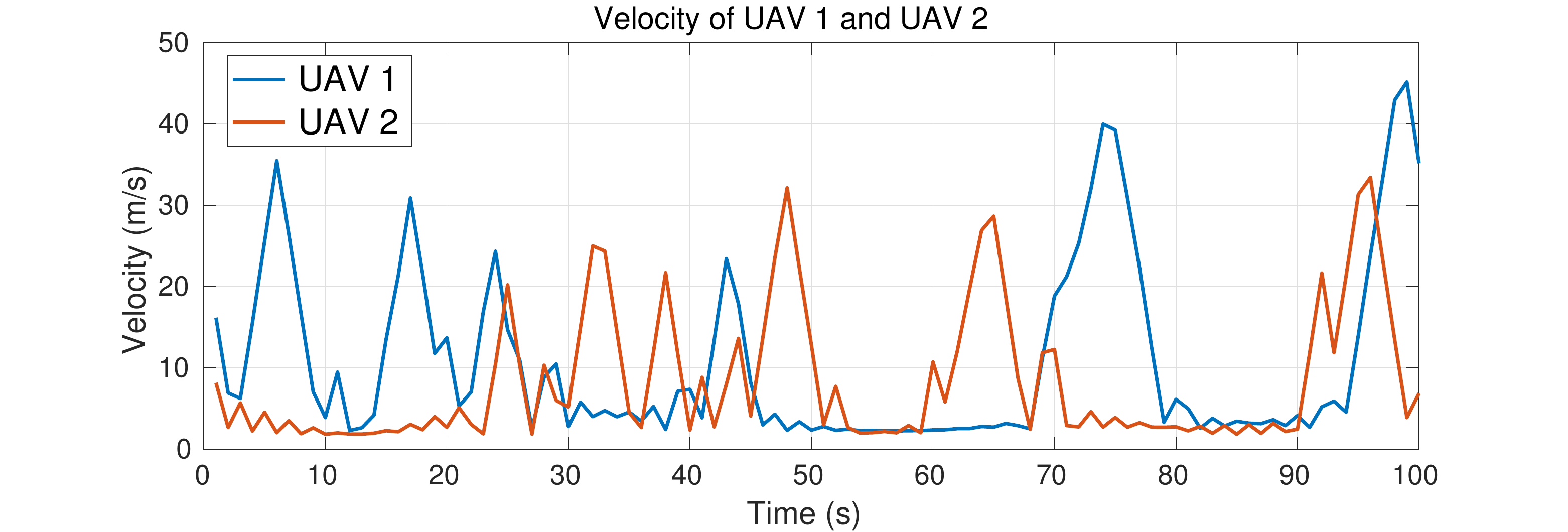}  
  \caption{UAV's velocity for $T=100$ s}
  \label{fig:sub-third}
\end{subfigure}
\caption{Optimal parameters for our proposed algorithm}
\label{fig:fig}
\end{figure}

 We note that, in the initial solution that we provided, the UAV 2 communicates only with two ground stations which are represented by the red dots in Figure \ref{fig:sub-first}. After applying our proposed algorithm, the UAV 2 extended its trajectory in order to serve a third ground station that was only served by the first UAV in the initial solution as shown in Figure \ref{fig:sub-first}.

We also note that every UAV tries to position itself near the ground station when there is a communication between them. This is explained by the fact that this helps them achieve better communication channel. For instance, all the UAVs tend to have a high velocity when they are moving from a ground station to another  and tend to have a minimum velocity when they are hovering next to a ground station as displayed in the Figure \ref{fig:sub-third}. However, the UAVs cannot maintain a fixed position during their flight due to the minimum velocity constraint and the constraint of energy. As a result, they continuously revolve around the ground stations as closely as possible with a minimum velocity in order to preserve good communication channels.\\
According to the Figure \ref{fig:sub-second}, we note that the transmission power of the different UAVs is almost always set at its maximum level $P_{max}$. This is explained by the fact that, in the resulting solution of our algorithm, the UAVs are quite distant from each other during their communications with the ground stations. As a result, our resulting solution could be assimilated to the combination of two distant subsystems in which every one of them is composed of one UAV. However, The impact of the power transmission would be more tangible if we adopted a system which is composed of a large number UAVs. In that case, the communication channel would be highly impacted by the interference at the level of users as there are a lot of UAVs that are close to each other emitting the signals at the same time.\\

\textbf{Remark:}
We can always consider working with a larger number of UAVs and ground stations. For instance, Figure 2 illustrates the optimal trajectories of UAVs in the case where we consider a system composed of 4 UAVs and 9 ground stations. 
\begin{figure}[h]
\centering
\includegraphics[scale=0.2]{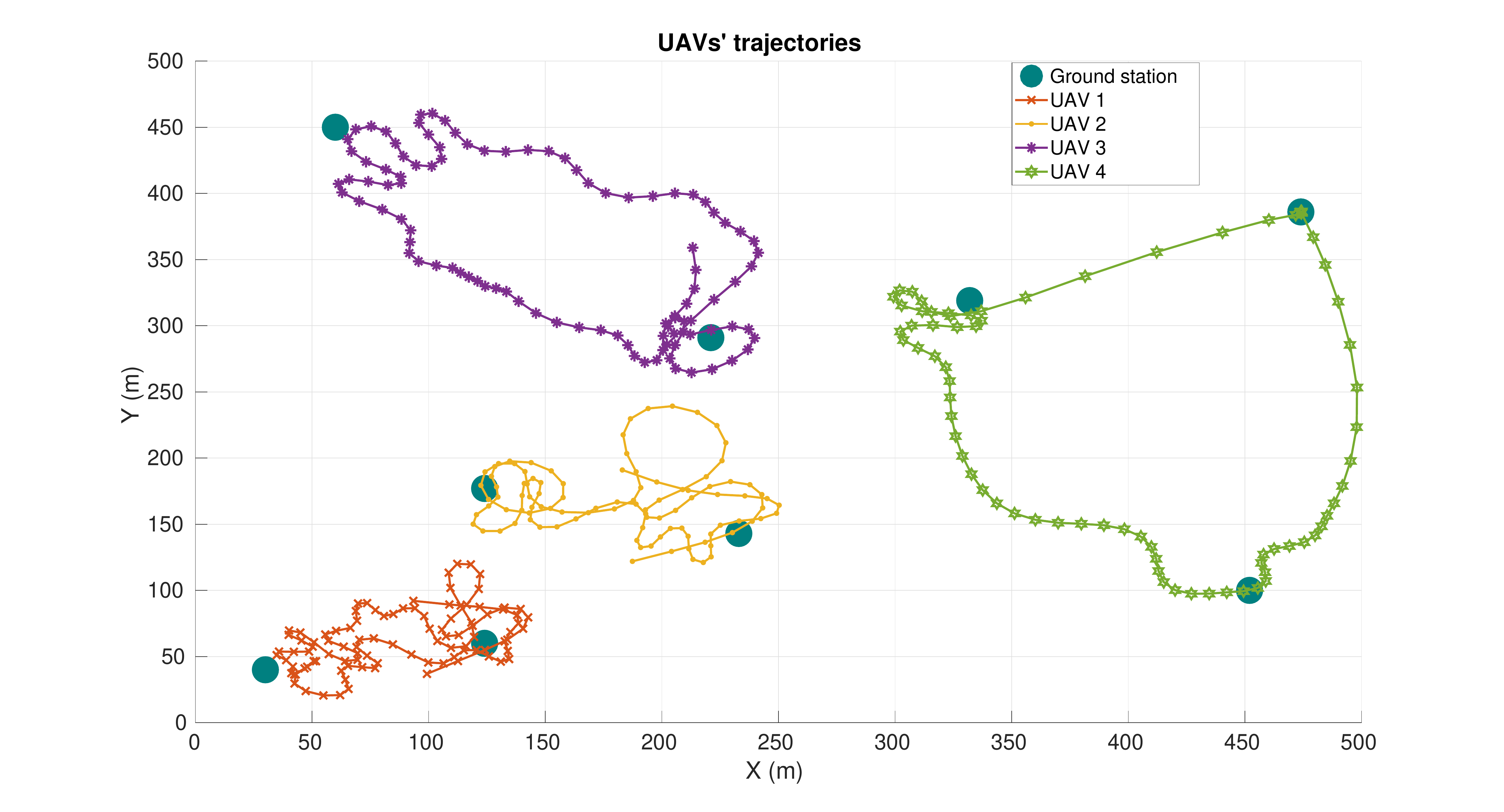}
\caption{UAVs' optimal trajectories for $T=100$ s.}
\label{fig4}
\end{figure}

\subsection{Convergence of our proposed Algorithm}
Let us study the convergence behavior of our algorithm. It can be observed from the Table \ref{Tab1} below that the max-min throughput increases in every iteration. We  also note that our algorithm converges quickly to a final solution within about 10 iterations.
\begin{table}[h]

\caption{Convergence Behavior}\label{Tab1}
\begin{center}
\begin{tabular}{c||c|c|c|c|c|c|c|c|c} 
\hline
$r$ &  1 & 2 & 3 & 4 & - & 8 & 9 & 10 & 11  \\ \hline
 $\Delta \mu^{lb}$ & 125 & 16.5 & 8 & 2.3 & - & 0.3 & 0.2 & 0.2 & 0.1  \\
\hline
\end{tabular}
\end{center}
\vspace{-0.3in}

\end{table}
\subsection{Performance Analysis}
Let us compare the performance of our proposed solution to a benchmark scheme with static access point, placed in the geometric center of the ground nodes. In fact, the baseline scheme allows us to achieve a performance of $R= 17$ Bps/Hz, which is much more negligible to the performance achieved by our solution which is $R= 208$ Bps/Hz.  This result strongly highlight the benefit brought by UAV mobility in such problems.  

 Additionally, let us compare the performance of our proposed solution to the initial trajectory solution. We note that our proposed solution allows us to achieve results that are significantly better than those obtained from the initial solution. Indeed, the max min rate has increased from $R= 70$ Bps/Hz to $R= 208$ Bps/Hz for a duration  $T=100$ s. This represents an improvement of nearly $150$ Bps/Hz in the overall performance.
\begin{table}[H]
\caption{Performance achieved by the different ground stations}\label{Tab2}
\begin{center}
\begin{tabular}{c||c|c|c|c|c|c} 
\hline
GT No. &  1 & 2 & 3 & 4 & 5 & 6\\ \hline
Data rate & \multirow{2}{*}{209.1} & \multirow{2}{*}{208.6} & \multirow{2}{*}{209.4} & \multirow{2}{*}{210.1} & \multirow{2}{*}{208.8} & \multirow{2}{*}{208.4}  \\
(Bps/Hz) &  &  &  &  &  &   \\ \hline
Connection time & \multirow{2}{*}{34} & \multirow{2}{*}{31} & \multirow{2}{*}{31} & \multirow{2}{*}{32} & \multirow{2}{*}{36} & \multirow{2}{*}{31} \\
(sec) &  &  &  &  &  & \\
\hline
\end{tabular}
\end{center}
\end{table}
\par
We also note  from the Table \ref{Tab2} that the UAVs have planned their movements so that the users have the same duration of communication with them. Additionally, we note  that the ground stations have received roughly the same rate which is equal to $R =208$ Bps/Hz. Thus, we consider that the objective of achieving a fair performance among all users is achieved.

\subsection{Impact of the energy on the solution}

In this section, we  study the impact of the energy constraint on the performance of our system. So in the beginning,  we denote $E_{m}$ as the energy consumed by the UAVs in the first example provided in Section V for $T=100 s$. Next, we  solve our problem for different values of $E_{max}$ and compare the results that we obtain with the first solution provided in Section V.
 \begin{itemize}
  \item $E_{max}$=0.9 $E_{m}$ \\ 
  
      \begin{figure}[H]
\centering
\includegraphics[scale=0.28]{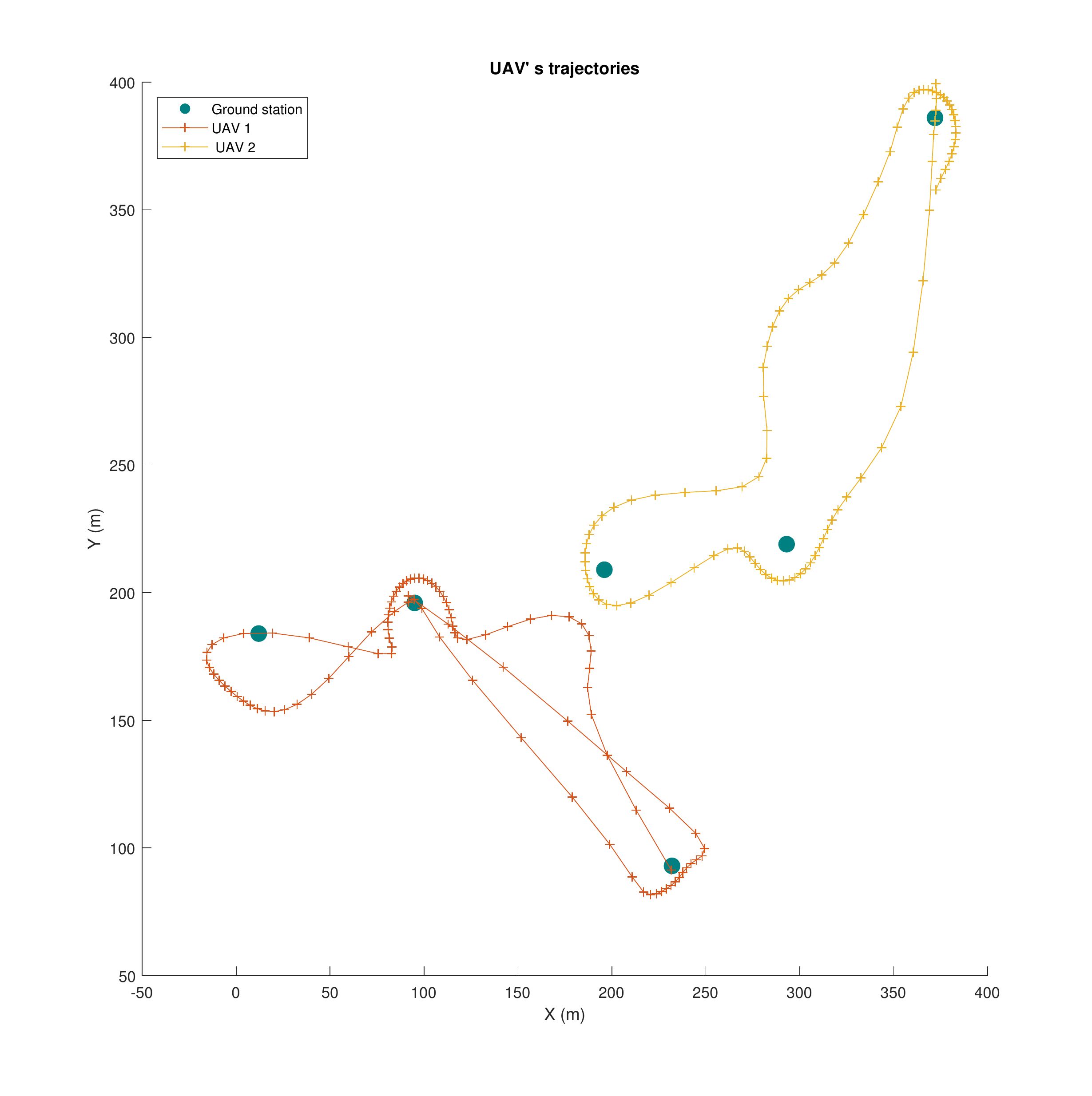}
\caption{UAVs' trajectories with constrained energy when $E_{max}=0.9 E_{m}$}\label{Fig:4}
\end{figure}

  We note that that the UAVs managed to find trajectories that are different from those we have in Figure \ref{Fig:4}. But, it succeeded to preserve the same max min rate $R=208$ bps/Hz. 

    \item $E_{max}$= 0.6 $E_{m}$\\
     \begin{figure}[H]
\centering
\includegraphics[scale=0.6]{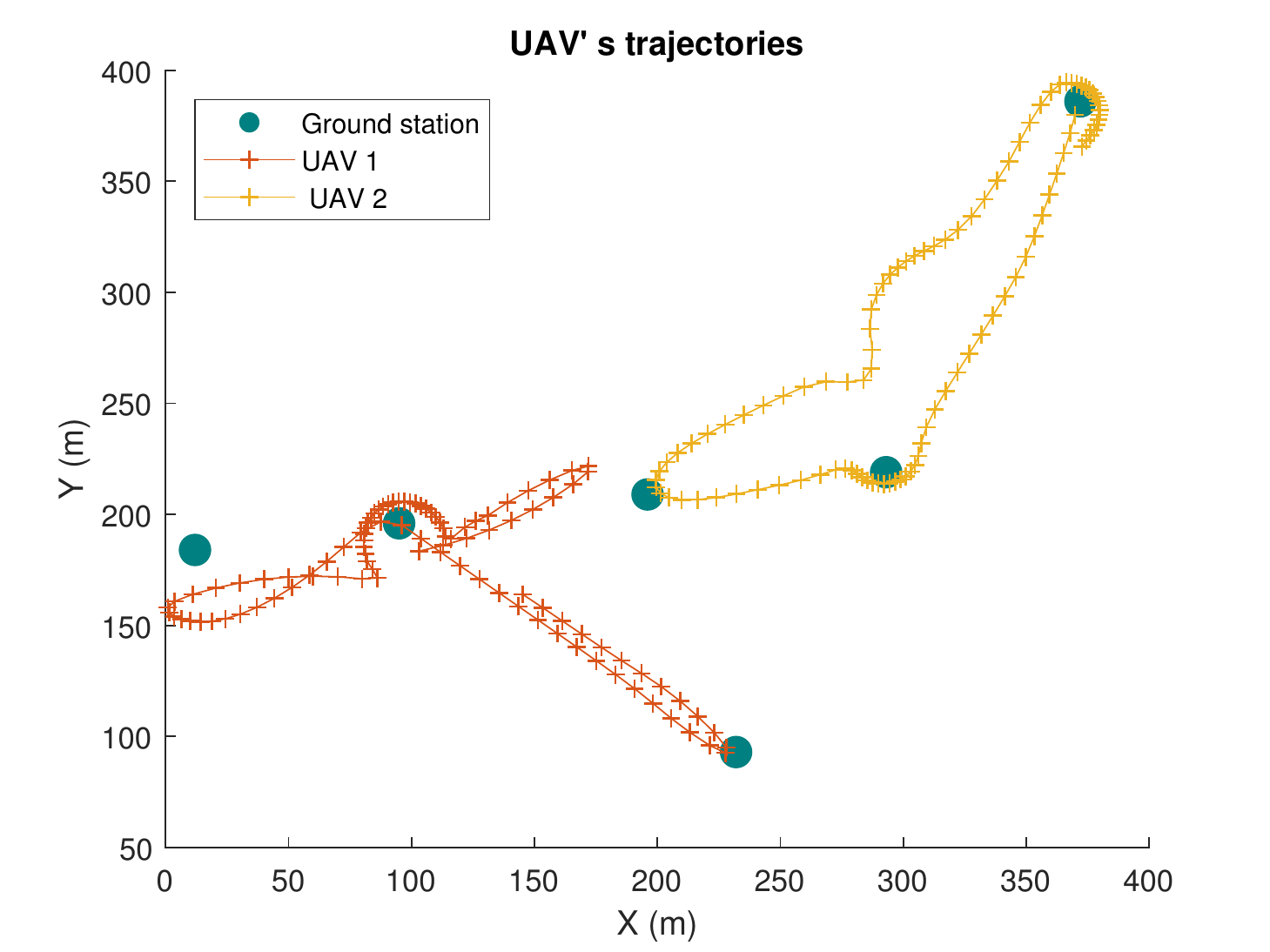}
\caption{UAVs' trajectory with constrained energy when $E_{max}=0.6 E_{m}$}\label{Fig:5}
\end{figure}

    Our algorithm succeeded to converge to a final solution which is represented in Figure \ref{Fig:5}. However, we note that the value of the max min rate has fallen drastically compared with the initial solution, from $R=208$ bps/Hz to $R= 183.6$ bps/Hz.
        \item $E_{max}$= 0.3 $E_{m}$\\
    We note that our algorithm failed to converge because it was not able to find a feasible set for our problem that is compliant with the different constraints of energy and velocity.
\end{itemize}

\section{Conclusion}
\label{conclusion}

In this letter, we have investigated the max-min fairness problem between users in wireless communication  with multiple UAVs. The main objective behind this problem was to design an efficient solution that jointly optimizes the UAVs' trajectories, the transmission power and the use scheduling that meet certain quality of service requirements. To this end, we proposed an effective iterative algorithm that regroups the BCD and the SCA techniques. Then, we demonstrated the effectiveness and performance of our algorithm compared to the baseline circular path planning scheme.
\appendices
\section{}
\hypertarget{Appendix A}{}
In this section, we will solve \hyperlink{P3}{\textbf{(P3)}} using the successive convex optimization techniques. Let us first start by the constraints (\ref{constEm}) and (\ref{a3}). The resulting sets of these constraints are not convex with respect to $\mathcal{V}$.  
Thereby, we start by introducing a slack variable $\lambda_{m}(n)$ and replacing it with $\| \mathbf{v}_{m}(n) \|$ 
. As a result, by defining  $\Lambda=\{\lambda_{m}(n)$ \hspace{0.1 cm} ,$\forall m,n$\} we obtain the following problem \hyperlink{P3.a}{\textbf{(P3.a)}} :
\begin{small}
\begin{eqnarray}
\hypertarget{P3.a}{}
& \max\limits_{\mu^{lb},\mathcal{B,Q,V,A},\Lambda} & \mu^{lb} \nonumber\\ 
& \textrm{s.t.} & (\ref{a5})-(\ref{amax}),\indent (\ref{a15}) \nonumber\\
&(\textbf{P3.a})& 0\leq \displaystyle\sum_{n=1}^{N} c_{1} \| \mathbf{v}_{m}(n) \| ^3  +\frac{c_{2}}{\lambda_{m}(n) }\left(1+\frac{\| \mathbf{a}_{m}(n) \|^2}{g }\right)+\frac{w}{2} (\| \mathbf{v}_{m}(N) \|^2-\| \mathbf{v}_{m}(0) \|^2)    \leq E_{max},   \label{a18}  \\
&& \lambda_{m}(n)\geq V_{min}, \indent \forall m,n \label{a16} \\
&& \| \lambda_{m}(n) \| ^2 \leq \| \mathbf{v}_{m}^r (n) \|^2,	\indent \forall m,n   \label{a17}
\end{eqnarray}
\end{small}
$\Lambda$ is now considered as a new decision variable set. Additionally, we have also two new constraints (\ref{a16}) and (\ref{a17}). It is worth noting that at the optimal solution of this new problem \hyperlink{P3.a}{\textbf{(P3.a)}}, we must have $\lambda_{m}(n)=\|\mathbf{v}_{m}(n)\|$. Otherwise, we can enlarge the feasible region corresponding to the constraint (\ref{a17}) by increasing $\lambda_{m}(n)$. Therefore, we can say with certainty that the problem \hyperlink{P3.a}{\textbf{(P3.a)}} is equivalent to the previous problem \hyperlink{P3}{\textbf{(P3)}}.
\par
In addition, with this new reformulation, the resulting set of the energy constraint (\ref{a18}) is now convex with respect to ($\mathcal{V,A},\Lambda$). The constraint (\ref{a16}) is also convex with respect to $\Lambda$. However, the new constraint (\ref{a17}) is non-convex with respect to $\mathcal{V}$.
\par
Next, As stated earlier, we are trying to solve \hyperlink{P3.a}{\textbf{(P3.a)}} approximately by solving iteratively a series of convex problems that represent a lower bound to this latter. So, as we want to maximize $\mu$, our approach consists into shrinking the feasible set of the non-convex constraints by making some approximations in order to make them convex.
\par Specifically, we apply the successive convex optimization. So, we define the given velocity of an UAV  in the $r^{th}$ iteration as $\mathcal{V}^r$ =\{$\mathbf{v}^{r}_{m}(n)$ \hspace{0.1 cm} ,$\forall m,n\}$ which represents the solution of $\mathcal{V}^r$ computed at the $(r-1)^{th}$ iteration. Then, we use the following lemma:
\begin{lemma}
\hrule
\vspace{2mm}
If $f$ is a convex (respectively concave) function with respect to a certain variable $x \in X$. Then $f$ is lower (respectively upper) bounded by its first order Taylor expansion at any point $x, y \in  X$. Thus: \begin{itemize}
    \item $f(x)\geq f(y)+f’(y)(x-y)$, \hspace{0.3cm} if $f$ is convex
    \item $f(x)\leq f(y)+f’(y)(x-y)$,\hspace{0.3cm} if  $f$ is concave
\end{itemize}
\vspace{2mm}
\hrule
\vspace{2mm}
\end{lemma}
Since $f(x)$= $\| x \|^2$ is convex, we apply the previous lemma by taking  $x=  \mathbf{v}_{m} (n) $ and \newline $y =  \mathbf{v}_{m}^r (n)  $ 
  
\par Consequently, we have the following result: \begin{equation*}
    	\| \mathbf{v}_{m} (n) \|^2 \geq \| \mathbf{v}_{m}^r (n) \|^2 + 2   \left(\mathbf{v}_{m}^{r}(n)\right)^T( \mathbf{v}_{m} (n)-\mathbf{v}_{m}^r (n)).	
\end{equation*}
							                
\par As a result, we define the new constraint 
\begin{equation*}
     \mathbf{\lambda}_{m} (n)^2 \leq \| \mathbf{v}_{m}^r (n) \|^2 + 2  \left(\mathbf{v}_{m}^{r}(n)\right)^T( \mathbf{v}_{m} (n)-\mathbf{v}_{m}^r (n)),	  \forall m,n
\end{equation*}
	
\par This new constraint is convex with respect to ($\mathcal{V},\Lambda$).

\par Let us consider now the constraint (\ref{a15}) which is non-convex with respect to $\mathcal{B}$. In order to deal with it, we first write the left-hand side of the constraint as a difference between two concave functions with respect to the auxiliary variable $\mathcal{B}$. So, we have: 

\begin{eqnarray}
    R_{k,m}(n)&=& \log_{2}\left(1+\frac{B_{k,m}(n)}{\sum_{\substack{
   j=1 \\
   j\neq m
  }} ^{M}
B_{k,j}(n)+ \sigma_{0} ^2}{}\right)\nonumber \\ 
&=&\log_{2}\left(\sum_{\substack{
   j=1
  }} ^{M}B_{k,j}(n)+\sigma_{0}^2\right) - \overline{R}_{k,m}(n),\nonumber 
\end{eqnarray}
 where  $\overline{R}_{k,m}(n)= \log_{2}\left(\sum_{\substack{
   j=1\\ j\neq m
  }} ^{M}B_{k,j}(n)+\sigma_{0}^2\right)$

  As the first and the second terms of $R_{k,m}(n)$ are concave, we cannot study the convexity of $R_{k,m}(n)$. To tackle this problem, we can use the successive convex optimization in order to approximate  $\Bar{R}_{k,m}$(n) by a more controllable function at a given local point. Specifically, we define  $\mathcal{B}^{r}$ =\{$B^{r}_{k,m}(n)$ \hspace{0.1 cm} $\forall k,m,n$\}  which represents the given auxiliary variable calculated in the $(r-1)^{th}$ iteration. 
 \par Since $f(x)= \log(x)$ is concave, we apply the previous lemma by taking \newline
  $x=\sum_{\substack{
   j=1\\ j\neq m
  }} ^{M}B_{k,j}(n)+\sigma_{0}^2$
    and $y = \sum_{\substack{
   j=1\\ j\neq m
  }} ^{M}B^r_{k,j}(n)+\sigma_{0}^2$
\\

We obtain the following result:
\begin{small}
$\overline{R}_{k,m}(n) \leq \overline{R}_{k,m}^{upper}(n)=\sum_{\substack{
   j=1 \\
   j\neq m
  }} ^{M}
A_{k,j}(n)(B_{k,j}(n)-B_{k,j}^r(n)) +C_{k,j}(n){} $ \\
\end{small}
where the constants $A_{k,j}(n)$ and $C_{k,j}(n)$ are given by \begin{eqnarray}
   A_{k,j}(n)=&\frac{\log_{2}(e)}{\sum_{\substack{
   j=1\\ j\neq m
  }} ^{M}B_{k,j}^r(n)+\sigma_{0}^2} \nonumber\\
  C_{k,j}(n)=&\log_{2}\left(\sum_{\substack{
   j=1\\ j\neq m
  }} ^{M}B_{k,j}^r(n)+\sigma_{0}^2\right) \nonumber
\end{eqnarray}
As a result, for a given trajectory $B^{r}$ defined in the $r^{th}$ iteration, the non-convex constraint (\ref{a1}) can be approximated by:
\begin{eqnarray}
\frac{1}{N} \displaystyle\sum_{n=0}^{N} \displaystyle\sum_{m=1}^{M}\left(\alpha_{k,m}(n) \log_{2}(\sum_{\substack{
   j=1
  }} ^{M}B_{k,j}(n)+\sigma_{0}^2) - \overline{R}_{k,m}^{upper}(n)\right)\geq \mu \nonumber
\end{eqnarray}
\par Now, Let us consider the constraint (\ref{a15}). The resulting set of this constraint is not convex with respect to $\mathcal{Q}$. So, as explained in the beginning of this section, we need to find a lower bound to the right-hand side of this constraint which is convex.
\par As shown in \hyperlink{Appendix D}{\textbf{(Appendix D)}}, the function  $B_{k,m,max}(n)=\frac{\beta_{0}P_{max}}{H^2+ \| \mathbf{q}_{m} (n) - \mathbf{w}_{k} \| ^2}{} $ is a concave surrogate function. As a result, after we define $\mathcal{Q}^{r}$ =$\{Q^{r}_{m}(n)$ \hspace{0.1 cm} ,$\forall m,n\}$  which represents the given trajectory calculated in the $(r-1)^{th}$ iteration, we have the following results:
\begin{eqnarray}
B_{k,m, max}(n)\geq P_{max}\beta_{0}\left(- \frac{\|\mathbf{q}_{m} (n) - \mathbf{w}_{k} \| ^2}{H^4}+D_{k,m}(n) (\mathbf{q}_{m} (n) - \mathbf{w}_{k})^T(\mathbf{q}_{m}^r (n) - \mathbf{w}_{k})+F_{k,m}(n) \right), \nonumber
\end{eqnarray}
where the constants $D_{k,m}(n)$ and $F_{k,m}(n)$ are given by \begin{small} \begin{eqnarray}
    D_{k,m}(n)\!\!&\!\!=\!\!&\!\!2\left(\frac{1}{H^4}-\frac{1}{(\| \mathbf{q}_{m}^r (n) - \mathbf{w}_{k} \| ^2+H^2)^2}\right)\nonumber\\
    F_{k,m}(n)\!\!&\!\!=\!\!&\!\!\frac{1}{\|\mathbf{q}_{m}^r (n) - \mathbf{w}_{k} \| ^2+H^2}+ \frac{2\| \mathbf{q}_{m}^r (n) - \mathbf{w}_{k} \| ^2}{(\| \mathbf{q}_{m}^r (n) - \mathbf{w}_{k} \| ^2+H^2)^2}\nonumber\\
    \!\!&\!\!-\!\!&\!\! \frac{\| \mathbf{q}_{m}^r (n) - \mathbf{w}_{k} \| ^2}{H^4} \nonumber
    \end{eqnarray}
    \end{small}
\par As a result, for a given trajectory $\mathcal{Q}^{r}$ defined in the $r^{th}$ iteration, the non-convex constraint (\ref{a15}) can be approximated by:
\begin{eqnarray}
     0\leq B_{k,m}(n)&\leq&  P_{max}\beta_{0}(- \frac{\| \mathbf{q}_{m} (n) - \mathbf{w}_{k} \| ^2}{H^4}\nonumber\\
     &+&D_{k,m}(n)(\mathbf{q}_{m} (n) - \mathbf{w}_{k})^T(\mathbf{q}_{m}^r (n) - \mathbf{w}_{k})\nonumber\\
     &+&F_{k,m}(n) ) \nonumber
\end{eqnarray}
\par Finally, for a given trajectory, velocity and the auxiliary variable ($\mathcal{Q}^r,\mathcal{V}^r,\mathcal{B}^r$), we obtain the problem (\textbf{P4}).
\par Thanks to the lower bounds that we adopted in the constraints (\ref{a1}), (\ref{a15}) and (\ref{a17}), the resulting set of all the constraints of the problem \hyperlink{P4}{\textbf{(P4)}} are convex. As a result, the problem \hyperlink{P4}{\textbf{(P4)}} is now considered as a convex optimization problem that can be solved efficiently using some predefined optimization solvers.

\section{ Proof of the convergence and complexity of our Algorithm}

\hypertarget{Appendix B}{}
\subsection{Proof of convergence}
In order to prove that our algorithm converges, we first start by defining the objective values of our main problem \hyperlink{P2}{\textbf{(P2)}} $\mu$ as well as the sub-problems \hyperlink{P2.1}{\textbf{(P2.1)}} and \hyperlink{P4}{\textbf{(P4)}} at the $r^{th}$ iteration as $\mu^r(\mathcal{S})=\mu^r$  and $\mu^{lb,r}( \mathcal{A,Q,V,S,B})=\mu^{lb,r}$   , respectively. \\

Since \hyperlink{P2.1}{\textbf{(P2.1)}} is defined as a sub-problem of \hyperlink{P2}{\textbf{(P2)}} for fixed $(\mathcal{Q}^{r},\mathcal{B}^{r},\mathcal{V}^{r})$, then when we optimize \hyperlink{P2.1}{\textbf{(P2.1)}} over the user scheduling the objective value $\mu$ of the global problem \hyperlink{P2}{\textbf{(P2)}} is also maximized for a given $(\mathcal{Q}^{r},\mathcal{B}^{r},\mathcal{V}^{r})$. Then:
\begin{eqnarray}
\mu(\mathcal{S}^{r}, \mathcal{Q}^{r},\mathcal{V}^{r}, \mathcal{B}^{r}) \leq \mu(\mathcal{S}^{r+1}; \mathcal{Q}^{r},\mathcal{V}^{r}, \mathcal{B}^{r}) \label{i}
\end{eqnarray}

Second, for given $(\mathcal{S}^{r+1}, \mathcal{Q}^{r},\mathcal{V}^{r}, \mathcal{B}^{r})$ in step \hyperlink{2.b}{\textbf{2.b)}} of our algorithm , we can express the following relationships:
\begin{eqnarray}
      \mu(S^{r+1}; Q^{r},B^{r},V^{r}) &\stackrel{(a)}{=}& \mu^{lb,r}(\mathcal{S}^{r+1}, \mathcal{Q}^{r},\mathcal{V}^{r}, \mathcal{B}^{r}) \nonumber\\
 &\overset{(b)}{\le}&\mu^{lb,r}(\mathcal{S}^{r+1}, \mathcal{Q}^{r+1},\mathcal{V}^{r+1}, \mathcal{B}^{r+1})\nonumber\\
&\overset{(c)}{\le}&\mu(\mathcal{S}^{r+1}, \mathcal{Q}^{r+1},\mathcal{V}^{r+1}, \mathcal{B}^{r+1})  \label{ii} 
\end{eqnarray}

where the first equation \textbf{(a)} holds since the surrogate functions in (\ref{a15}) as well as the first-order Taylor expansions in (\ref{a1}) and (\ref{a17}) are tight at given local points. In other words, for given ($\mathcal{Q}^{r},\mathcal{V}^{r}, \mathcal{B}^{r})$, we can consider that problem \hyperlink{P4}{\textbf{(P4)}} has the same objective value as that of problem  \hyperlink{P2.2}{\textbf{(P2.2)}}.\\

The second inequality (b) is derived from the fact that ($\mathcal{Q}^{r+1},\mathcal{V}^{r+1}, \mathcal{B}^{r+1}$) represent the optimal solutions of problem \hyperlink{P4}{\textbf{(P4)}} for given ($\mathcal{S}^{r+1},\mathcal{Q}^{r},\mathcal{V}^{r}, \mathcal{B}^{r}$) and thanks to the non-decreasing propriety of the objective function in  this problem   (because we are maximizing $\mu$), this inequality holds.\\

The third inequality (c) holds since the objective function of the approximation problem \hyperlink{P4}{\textbf{(P4)}} is considered as lower bound of that of the original sub-problem  \hyperlink{P2.2}{\textbf{(P2.2)}}.\\

Based on  (\ref{i}) and (\ref{ii}), we have the following result: \\

$\mu(\mathcal{S}^{r}, \mathcal{Q}^{r},\mathcal{V}^{r}, \mathcal{B}^{r})\leq \mu(\mathcal{S}^{r+1}, \mathcal{Q}^{r+1},\mathcal{V}^{r+1}, \mathcal{B}^{r+1})$\\

As a result, we can affirm that, in each iteration, the objective value   $\mu$ of our main problem \hyperlink{P2}{\textbf{(P2)}} increases or at least remains unchanged. Additionally, since the objective value is upper bounded, the convergence of our algorithm is thus proved.
\subsection{Proof of complexity of our algorithm}

In order to determine the complexity of our algorithm, we should determine the complexity of both subproblems (P2.1) and (P4) defined in section IV of our paper.

First, problem (P2.1) is a standard linear problem that depends only on the user-UAV association parameters. As explained in \cite{compl}, the complexity of such problems is of order $\Theta\left(a^{2} b\right)$, where $a$ is the number of variables and $b$ is the number of constraints. Following this, it is easy to show that the complexity of (P2.1) is of order $\Theta\left( (KMN+MN+KN+K)(KM)^{2}  \right)$.

Second, the problem (P4) is defined as follows:

\begin{small}
\begin{eqnarray}
\!\!\!\!\!\!\!\!\!\!\!\!\!\!&\!\!\!\!\!\!\!\!\!\!\!\! \!\!\!\!\!\max\limits_{\mu^{lb},\mathcal{B,Q,V,A},\Lambda} & \mu^{lb} \nonumber\label{a1}\\ 
&\!\!\!\!\!\!\!\!\!\!\!\!\!\!\! \textrm{s.t.} \!\!\!\!&\!\!\!\!\!\!\!\! \mathbf{v}_{m}(n+1) = \mathbf{v}_{m}(n) + \mathbf{a}_{m}(n) \delta t,\\
&&\!\!\!\!\!\!\!\!\mathbf{q}_{m}(n+1) = \mathbf{q}_{m}(n) + \mathbf{v}_{m}(n) \delta t + \frac{1}{2}\mathbf{a}_{m}(n) \delta  t^2,\\
&&\!\!\!\!\!\!\!\!\| \mathbf{v}_{m}(n) \| \leq V_{max} ,\hspace{0.5cm}\forall m,n,\\
&&\!\!\!\!\!\!\!\!\| \mathbf{a}_{m}(n) \| \leq a_{max} ,\hspace{0.5cm}\forall m,n,\\
&&\!\!\!\!\!\!\!\!  \frac{1}{N} \displaystyle\sum_{n=0}^{N} \displaystyle\sum_{m=1}^{M}\alpha_{k,m}(n)\log_{2}\left(\sum_{\substack{
   j=1
  }} ^{M}B_{k,j}(n)+\sigma_{0}^2\right) - \overline{R}_{k,m}^{upper}(n))\geq \mu^{lb}, \indent  \forall k \\
&&\!\!\!\!\!\!\!\! 0\!\leq\! B_{k,m}(n)\!\leq \! P_{max}\beta_{0}\!\left(\!- \frac{\| \mathbf{q}_{m}(n)\! - \!\mathbf{w}_{k}\| ^2}{H^4}\!+\!D_{k,m}(n)(\mathbf{q}_{m} (n) \!-\! \mathbf{w}_{k})^T(\mathbf{q}_{m}^r (n) \!-\! \mathbf{w}_{k})\!+\!F_{k,m}(n)\!\!\right)\!, \indent  \forall k,m,n  \nonumber \\
&&\\
&&\!\!\!\!\!\!\!\! 0\leq \displaystyle\sum_{n=1}^{N} c_{1} \| \mathbf{v}_{m}(n) \| ^3  +\frac{c_{2}}{\lambda_{m}(n) }\left(1+\frac{\| \mathbf{a}_{m}(n) \|^2}{g }\right)+\frac{w}{2} (\| \mathbf{v}_{m}(N) \|^2-\| \mathbf{v}_{m}(0) \|^2)    \leq E_{max},  \indent \forall m  \\
&&\!\!\!\!\!\!\!\! \lambda_{m}(n)\geq V_{min}, \indent \forall m,n  \\
&&\!\!\!\!\!\!\!\! \lambda_{m}(n) ^2 \leq \| \mathbf{v}_{m}^r (n) \|^2 + 2(\mathbf{v}_{m}^{r}(n))^T\left( \mathbf{v}_{m}(n)-\mathbf{v}_{m}^r (n)\right),	\indent \forall m,n   \\
\!\!\!\!&\!\!\!\!\!\!\!\!&\!\!\!\!\!\!\!\!\!\!\!\!\!\!\!\!\!\!\!\!\!\! {\color{blue}\|\mathbf{q}_{m}(n)\!-\!\mathbf{q}_{j}(n) \|^2 \!\leq\! \| \mathbf{q}_{m}^r (n)\!-\!\mathbf{q}_{j}^r (n) \|^2 \!+\! 2(\mathbf{q}_{m}^r (n)\!-\!\mathbf{q}_{j}^r (n))^T\!\left( \mathbf{q}_{m}^r (n)\!-\!\mathbf{q}_{j} (n)\!-\!\mathbf{q}_{m} (n)\!+\!\mathbf{q}_{j}^r (n)\right),	\indent \forall m,n,j\neq m } \nonumber\\
&&\label{a10}
\end{eqnarray}
\end{small}

Problem (P4) is a convex problem. As stated in [1], the complexity of (P4) is of order $\Theta\left( max(a^{3} , a^{2} b, F )\right)$ where $a$ is the number of variables, $b$ is the number of constraints and $F$ is the cost of evaluating the first and second derivatives of the objective and constraint functions.

\begin{itemize}
\item First, it can be easily proved that $a= KMN+7MN+1$.

\item Second, we can compute $b$ by summing the number of constraints in the equations (\ref{a1})-(\ref{a10}) which are represented in the following table:
\hspace{1in}\begin{table}[h]
\begin{center}
\begin{tabular}{c||c|c|c|c|c|c|c|c|c|c|c|c} 
\hline
Eq. &   (1) & (2) & (3) & (4) & (5) & (6) & (7) & (8) & (9) & (10)   \\ \hline
$b$ &  $M(N-1)$ & $M(N-1)$ & $MN$ & $MN$ & $K$ & $KMN$ & $M$ & $MN$ & $MN$ &  $\frac{ M(M-1)N}{2}$    \\
\hline
\end{tabular}
\end{center}
\end{table}\\
Thus, we can totally calculate the number of constraints as $b= KMN+4MN+2M(N-1)+K+M+\frac{M(M-1)N}{2} $.

\normalsize
\item Since  $F$ is the cost of evaluating the first and second derivatives of the objective and constraint functions, a simple approach consists into computing the first and second derivatives of  the equations (\ref{a1})-(\ref{a10})  with respect to each of the optimization variables ($\mu^{lb},\mathcal{B,Q,V,A},\Lambda$) separately and then summing the number of operations required to compute every single one of these quantities. By doing so, we get:
$F\approx122 MN+(4M+9)KMN$.
\end{itemize}

It is clear that $a^2b$ is greater than $a^{3}$ and $F$. Thus, the complexity of (P4) is
 \begin{align*}
 \small{\Theta \left((KMN+7MN+1)^{2}\left(KMN+4MN+2M(N-1)+K+M+\frac{M(M-1)N}{2}\right)\right).}
 \end{align*}

Finally, the complexity of one iteration of our algorithm is nothing but the sum of the complexities of the two subproblems which is 
\begin{eqnarray*}
&&\Theta\Bigg(\underbrace{(KMN+MN+KN+K)(KM)^{2}}_{(P2.1)} + \nonumber\\
&&\underbrace{(KMN+7MN+1)^{2}\left(KMN+4MN+2M(N-1)+K+M+\frac{M(M-1)N}{2}\right)}_{(P4)} \Bigg).
\end{eqnarray*}
If we assume that $K$, $M$ and $N$ are much larger than one ($\gg1$), then the complexity of 1 iteration of our algorithm can be reduced to the order of
\begin{eqnarray*}
\Theta\left(\underbrace{(KM)^{3}N}_{(P2.1)} +\underbrace{(KMN)^{2}\left(KMN+\frac{M^2N}{2}\right) }_{(P4)}\right).
\end{eqnarray*}

\section{ K-means clustering algorithm}

\subsection{Definition}
\hypertarget{Appendix C}{}
\par $K$-means clustering is an unsupervised learning algorithm which is mainly used in data mining and statistics. This iterative method aims to partition the data into $K$ clusters by allocating each point to the cluster with the nearest mean. In other words, this method clusters automatically similar data examples together.
\par  The intuition behind $K$-means is an iterative procedure that is explained in the following algorithm:
\vspace{0.2cm}
\hrule
\vspace{2.5mm}
\textbf{\textbf{Algorithm :} K-means Clustering Algorithm }  \vspace{2mm}
\hrule
\begin{enumerate}
  \item 	Centroids initialization.
  \item \textbf{Repeat}
  \begin{enumerate}
 \item 	Calculate distance between the points and the centroids.
 \item 	Assign the points to the closest cluster.
\item 	Update the position of the centroids.
\end{enumerate}
\item \textbf{Until Convergence}
\end{enumerate}
\hrule

\vspace{2mm}
As explained in the above algorithm, we first start by initializing the position of the centroids (points that represent the center of the clusters). In the second step, for each point, we calculate its distance to all the centroids and then assign it to the closest one. More formally, if we define $\mathcal{C}$ as the set of centroids, then each point $\mathbf{x}$ is attributed to a cluster based on:\par
\hspace{3cm} $\arg\min\limits_{\mathbf{c}_{i}\in \mathcal{C}} (\|\mathbf{c}_{i}-\mathbf{x}\|)$
\par In the third step, we update the position of the centroids by taking the mean of the locations of all the points assigned to that centroid’s cluster. More precisely, the new positions of the centroids are given by :
\begin{equation}
    \mathbf{c}_{i}=\frac{1}{|\mathcal{S}{i}|} \displaystyle\sum_{\mathbf{c}_{i}\in \mathcal{S}_{i}}^{}\mathbf{x}_{i}\nonumber,
\end{equation}
where $\mathcal{S}_{i}$ and $|\mathcal{S}_{i}|$ are respectively the set and the number of the ground stations that belongs to the $i^{th}$ cluster.
\par
 Finally, we repeat iteratively the second and third steps until there is no change in the position of centroids.
 \subsection{Application of the $K$-means Clustering Algorithm}
We apply the $K$-means clustering algorithm in order to divide the ground stations into $M$ clusters. First, we randomly initialize the position of the $M$ centroids. Then, as explained in the previous subsection, we first assign every ground station to the closest centroid and then we update the position of these centroids by taking the mean of the cluster. So, this process is repeated iteratively until convergence to a final solution. 
\par
Next, we attribute every cluster to a UAV $m$. So, every UAV $m$ will communicate only with the ground stations that are located in its given cluster. 
 \begin{figure}[H]
\centering
\includegraphics[scale=0.8]{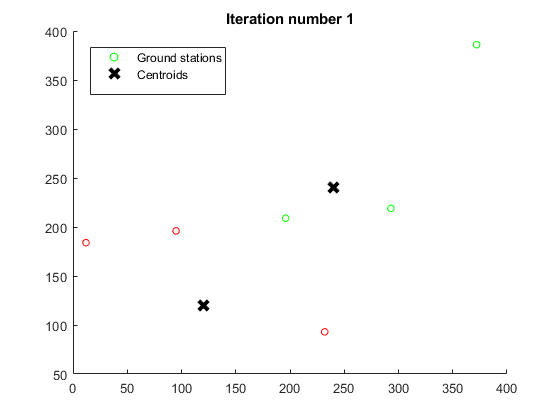}
\caption{Ground stations assignment in the first iteration}
\label{fig1}
\end{figure}
\begin{figure}[H]
\centering
\includegraphics[scale=0.75]{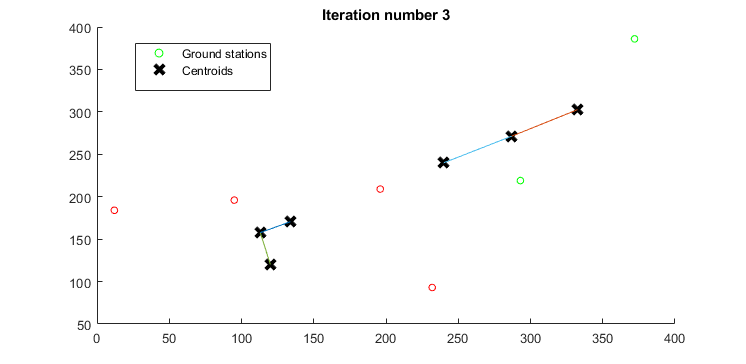}
\caption{Ground stations assignment in the third iteration}
\label{fig2}
\end{figure}
\par
 The figures \ref{fig1} and \ref{fig2} illustrate the iterative process that we used. So, in the end, we obtain two clusters of ground stations in which the red dots are assigned to one UAV  and the blue dots are assigned to the other UAV.
 \par Afterwards, we provide a circular trajectory as an initialization scheme for every UAV in which the centroids that we obtained from the $K$-mean clustering algorithm will represent the center of the circular trajectory of every UAV. Furthermore, the radius of this trajectory circles will represent the average distance from the centroid to its assigned ground stations which belong to the same cluster. In other words the radius $r_{m}$ is defined by:
\begin{equation*}
    r_{m}=\frac{1}{|\mathcal{S}_{m}|} \displaystyle\sum_{\mathbf{w}_{k}\in \mathcal{S}_{m}}^{} \|\mathbf{w}_{k}-\mathbf{c}_{m}\|    
\end{equation*}
 
 where $\mathcal{S}_{m}$ and $|\mathcal{S}_{m}|$ represent respectively the set and the number of ground stations that belong to the cluster m.\\
 \begin{figure}[H]
\centering
\includegraphics[scale=0.7]{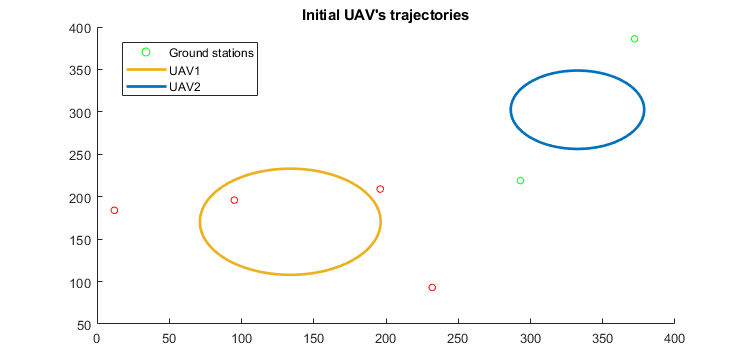}
\caption{Initial trajectories of the UAVs}
\end{figure}

\section{}
\hypertarget{Appendix D}{}
\label{Appendix}
Let us define $f(\mathbf{x})=\frac{1}{a \|\mathbf{x}\|^2 +b}$    ,where $a$ and $b$ are positive constants.
Then,  the function $g$ defined by:\begin{small}\begin{eqnarray}
  g(\mathbf{x}|\mathbf{x}_{0})\!\!&\!\!=\!\!&\!\!-\frac{a \|\mathbf{x}\|^2}{b^2}+2a\left(\frac{1}{b^2}-\frac{1}{(a \|\mathbf{x}_{0}\|^2 +b)^2}\right)\mathbf{x}^T\mathbf{x}_{0} \nonumber\\
   \!\!&\!\!+\!\!&\!\!\frac{1}{a \|\mathbf{x}_{0}\|^2 +b}+\frac{2a \|\mathbf{x}_{0}\|^2}{(a \|\mathbf{x}_{0}\|^2 +b)^2}-\frac{a \|\mathbf{x}_{0}\|^2 }{b^2}\ \forall \mathbf{x},\mathbf{x}_{0} \in R^2 \nonumber 
\end{eqnarray}  \end{small}is a concave surrogate function of $f$.
\par In order to prove this, we use the following lemma:
\begin{lemma}
\hrule
\vspace{2mm}
\par In order to prove that a function $g(\mathbf{x}|\mathbf{x}_{0})$ is a concave surrogate function of $f(\mathbf{x})$, $g(\mathbf{x}|\mathbf{x}_0)$  needs to satisfy the following conditions 
\begin{align*}  	
f(\mathbf{x})&= g(\mathbf{x}|\mathbf{x}) &,\forall \mathbf{x} \nonumber \\
  	\nabla f(\mathbf{x})&=\nabla g(\mathbf{x}|\mathbf{x})&,\forall \mathbf{x}\nonumber \\
	g(\mathbf{x}|x_{0})&\leq f(\mathbf{x})&,\forall \mathbf{x} \nonumber
\end{align*}
\hrule
\vspace{1mm}
\end{lemma}
\begin{IEEEproof}
\par
    \hypertarget{1}{\textbf{1)}}	It is easy to say that $f(\mathbf{x})= g(\mathbf{x}|\mathbf{x}), \hspace{1cm} \forall \mathbf{x}$\\
   \par \hypertarget{2}{\textbf{2)}} We first calculate the gradient of $f$ at any given point. So, we have:\\ $\nabla f(\mathbf{x}) = \frac{-2a}{(a \|\mathbf{x}\|^2 +b)^2}\mathbf{x}, \hspace{1 cm} \forall \mathbf{x} $\\Then we compute the gradient of $g$ with respect to $\mathbf{x}$. We obtain: \begin{small}
  \begin{align}
         \nabla g(\mathbf{x}|\mathbf{x}_0) = -\frac{2a\mathbf{x}}{b^2}+2a\left(\frac{1}{b^2}-\frac{1}{(a\|\mathbf{x}_{0}\|^2+b)^2}\right)\mathbf{x}_{0}, \hspace{0.2 cm} \forall \mathbf{x} \nonumber
    \end{align} \end{small}
     Thus, it can be easily shown that:\\ \hspace*{1cm} $\nabla g(\mathbf{x}|\mathbf{x}) =\nabla f(\mathbf{x}), \hspace{0.1 cm} \forall \mathbf{x} $\par
    \hypertarget{3}{\textbf{3)}} 	Let us show that: $g(\mathbf{x}|\mathbf{x}_{0}) \leq f(\mathbf{x}), \hspace{ 0.1cm} \forall \mathbf{x}$\\
    First we consider the mean value form of the Taylor series of the function $f(\mathbf{x}) - g(\mathbf{x}|\mathbf{x}_{0})$ at given point $\mathbf{x}_{0}$. Then we have:
    \begin{small}
   \begin{eqnarray}
 f(\mathbf{x}) - g(\mathbf{x}|\mathbf{x}_{0})&=&(f(\mathbf{x}_{0}) - g(\mathbf{x}_{0}|\mathbf{x}_{0}))\nonumber\\
 &+&(\nabla f(\mathbf{x}_{0}) - \nabla g(\mathbf{x}_{0}|\mathbf{x}_{0}))^T(\mathbf{x}-\mathbf{x}_{0})\nonumber\\
 &+&\frac{1}{2}(\mathbf{x}-\mathbf{x}_{0})^T(\nabla ^2 f(\mathbf{\epsilon}) - \nabla^2 g(\mathbf{\epsilon}|\mathbf{x}_{0}))\left(\mathbf{x}-\mathbf{x}_{0}\right)\nonumber ,
    \end{eqnarray}
     \end{small}where $\mathbf{\epsilon}$ is between $\mathbf{x}$ and $\mathbf{x}_{0}$.
     \par
    Since we proved in 1) and 2) that $f(\mathbf{x}) = g(\mathbf{x}|\mathbf{x})$ and  $\nabla g(\mathbf{x}|\mathbf{x}) =\nabla f(\mathbf{x})$, we can say that:\\
     $f(\mathbf{x}) - g(\mathbf{x}|\mathbf{x}_{0})=\frac{1}{2}(\mathbf{x}-\mathbf{x}_{0})^T(\nabla ^2 f(\mathbf{\epsilon}) - \nabla^2 g(\mathbf{\epsilon}|\mathbf{x}_{0}))(\mathbf{x}-\mathbf{x}_{0})$\par
     Additionally, it can be easily shown that :\begin{itemize}

     \item  $\nabla ^2 f(\mathbf{x})= \frac{-2a}{(a \|\mathbf{x}\|^2 +b)^2}I+\frac{8a^2}{(a \|\mathbf{x}\|^2 +b)^3}$\\
      \item $\nabla ^2 g(\mathbf{x}|\mathbf{x}_{0})= \frac{-2a}{b^2}I$ \end{itemize}
\par 
 We note that the eigenvalues of $\nabla ^2 f(\mathbf{\epsilon})$ are greater than the eigenvalues of $\nabla ^2 g(\mathbf{\epsilon}|\mathbf{x}_{0})$. Thus $\nabla ^2 f(\mathbf{\epsilon})-\nabla ^2 g(\mathbf{\epsilon}|\mathbf{x}_{0})$ is a positive semi definite matrix.\\
     
      As a result, $\frac{1}{2}(\mathbf{x}-\mathbf{x}_{0})^T(\nabla ^2 f(\mathbf{\epsilon}) - \nabla^2 g(\mathbf{\epsilon}|\mathbf{x}_{0}))(\mathbf{x}-\mathbf{x}_{0})\geq 0$\\
      
      Consequently , $g(\mathbf{x}|\mathbf{x}_{0}) \leq f(\mathbf{x}), \hspace{1 cm} \forall \mathbf{x}$ 
      \end{IEEEproof}
    Based on \hyperlink{1}{\textbf{1)}}, \hyperlink{2}{\textbf{2)}} and \hyperlink{3}{\textbf{3)}}, $g(\mathbf{x}|\mathbf{x}_{0})$ is a concave surrogate function of $f(\mathbf{x})$.\\
    
    By substituting $\mathbf{x} = \mathbf{q}_{m}(n)-\mathbf{w}_{k}$, $\mathbf{x}_{0} = \mathbf{q}_{m}^r(n) -\mathbf{w}_{k}$, $b = H^2$, and $a = 1$ and multiplying $f(\mathbf{x})$ and
$g(\mathbf{x}|\mathbf{x}_{0})$  by  $P_{max}\beta _{0}$  in the third condition of the lemma, we obtain the following result:
$ P_{max}\beta_{0}\left(- \frac{\| \mathbf{q}_{m}(n)-\mathbf{w}_{k} \| ^2}{H^4}+D_{k,m}(n) (\mathbf{q}_{m}(n)-\mathbf{w}_{k})^T(\mathbf{q}_{m}^r(n) -\mathbf{w}_{k})+F_{k,m}(n)\right) \leq  \frac{P_{max}\beta_{0}}{H^2+ \| \mathbf{q}_{m}(t) - \mathbf{w}_{k}\| ^2}$\\

where the constants $D_{k,m}(n)$ and $F_{k,m}(n)$ are given by: 
\begin{small}
\begin{eqnarray}
    D_{k,m}(n)\!\!&\!\!=\!\!&\!\!2\left(\frac{1}{H^4}-\frac{1}{(\| \mathbf{q}_{m}^r (n) - \mathbf{w}_{k} \| ^2+H^2)^2}\right)\nonumber\\
    F_{k,m}(n)\!\!&\!\!=\!\!&\!\!\frac{1}{\|\mathbf{q}_{m}^r (n) - \mathbf{w}_{k} \| ^2+H^2}+ \frac{2\| \mathbf{q}_{m}^r (n) - \mathbf{w}_{k} \| ^2}{(\| \mathbf{q}_{m}^r (n) - \mathbf{w}_{k} \| ^2+H^2)^2}\nonumber\\
    \!\!&\!\!-\!\!&\!\! \frac{\| \mathbf{q}_{m}^r (n) - \mathbf{w}_{k} \| ^2}{H^4} \nonumber
    \end{eqnarray}
    \end{small}

\bibliographystyle{IEEEtran}
\bibliography{References}
\end{document}